# Impedance budget and instability estimation of the HLS-Ⅱ storage ring*

Zhang Qing-kun(张青鹍)[1]   Wang Lin(王琳)[2]   Li Weimin(李为民)[3]   Gao Weiwei(高巍巍)

NSRL, School of Nuclear Science and Technology,
University of Science and Technology of China, Hefei 230029, P. R. China

**Abstract:** The upgrade project of Hefei Light Source storage ring is under way. In this paper, the wake fields of new designed vacuum chambers have been simulated by CST code, and then broadband impedances were obtained by FFT. Together with resistive wall wake fields, broadband impedance model for storage ring was established. Using theoretical formula, longitudinal and transverse single bunch instabilities were discussed. With carefully-designed vacuum chamber, the threshold of beam instabilities was higher than beam current goal.

**Key word**: wake fields; broadband impedance; CST; single bunch instabilities

**PACS:** 29.20.D-, 29.27.Bd

## 1   Introduction

The Hefei Light Source (HLS) is a dedicated VUV and soft X-ray light source, where beam energy is 800 MeV and circumference of storage ring is 66.13m. To improve quality of synchrotron radiation, an important Upgrade Project of HLS is undergoing, where storage ring will be reconstructed. The main parameters of HLS-II storage ring are listed in table 1. Except for lower beam emittance, the beam intensity will be higher than before, which is essential to high synchrotron radiation flux and brightness[1–2].

Geometric impedance-driven beam collective effects are the important considerations reaching higher beam intensity. Compared with old version of storage ring, significant efforts were made in new designed vacuum chambers.

In this paper, short range wake fields of various vacuum chamber in HLS storage ring were calculated using CST code, therefore broadband impedance model was established by FFT. Based on obtained impedance model, usual single bunch instability thresholds, such as microwave instability, parasitic energy loss, transverse model coupling instability, were estimated by classical formula [3–4].

## 2   Short review of Wakefield and impedance

When a charged particle is travelling through the storage ring, if the wall of the beam pipe is not perfectly conducting or contains discontinuities, the movement of

Table.1 Parameters of HLS-Ⅱ storage ring

| | |
|---|---|
| Beam Energy [MeV] | 800 |
| Circumference [m] | 66.13 |
| Momentum Compaction | 0.02 |
| beam emittance [nm-rad] | 40 |
| natural energy spread | 0.000472 |
| natural bunch length [ps] | 100 |
| Nominal rms bunch length [ps] | 50 |
| synchrotron turn [kHz] | 28.0 |
| tune | (4.4141, 3.2235) |
| average transverse beta function [m] | 8.50/5.25 |
| Energy lost per turn [keV] | 16.73 |
| Main rf frequency [MHz] | 204 |
| Beam current $I_{dc}$ [mA] | 300 |
| Damping time $\tau_x$ / $\tau_y$ / $\tau_z$ [ms] | 23/22/10 |

the image charges will be slowed down, thus leaving electromagnetic fields behind. The fields behind the particle are called wake fields, which are important because they influence the motion of the later particles

To determine the motion of the charged particles in the storage ring, we should study these fields. So the wake functions are defined as integrals over the normalized forces due to the EM fields excited in a structure by a point charge $q$ and evaluated at a distance $s$ behind it. Since the magnetic force is perpendicular to the direction of the particle motion, so we obtained the longitudinal and transverse wake function by integrating over the EM

* Supported by the Natural Science Foundation of China (11175182 and 11175180)
    1) E-mail: zhangqk@mail.ustc.edu.cn
    2) E-mail: wanglin@ustc.edu.cn
    3) E-mail: lwm@ustc.edu.cn

force normalized by the charge $q$ and a horizontal offset $\xi$ [3–5]:

$$\begin{cases} W_{\square}^{'} = -\dfrac{1}{q}\displaystyle\int_{-\infty}^{\infty} dz E_z \\ W_{\perp} = \dfrac{1}{q\xi}\displaystyle\int_{-\infty}^{\infty} dz \left(\vec{E} + \vec{v}\times\vec{B}\right)_{\perp} \end{cases} \quad (1)$$

However, if the exciting charge is a bunch of particles of finite length, and the distance to the test particle $s$ is measured from the centre of the bunch. Thus, the wake function is the wake potentials of a delta-function distribution. For a Gaussian bunch, the wake potential can in principle be found from the convolution of the wake function with the normalized line density $\lambda(\tau - t)$.

Actually, we often use coupling impedance for analytical study of the beam instabilities, which is the Fourier transformation of the wake function. Thus we can identify the coupling impedance of the vacuum chamber as:

$$\begin{cases} Z_{\square}(\omega) = \displaystyle\int_{-\infty}^{\infty} d\tau W_{\square}(\tau)\exp(-j\omega\tau) \\ Z_{\perp}(\omega) = -\dfrac{j}{\beta}\displaystyle\int_{-\infty}^{\infty} d\tau W_{\perp}(\tau)\exp(-j\omega\tau) \end{cases} \quad (2)$$

We also use the reduced impedance for the longitudinal beam instabilities, which is a comparable quality for storage rings with different circumference. The reduced impedance is defined as the impedance divided by the harmonic number $n = \omega/\omega_0$.

Meanwhile, the longitudinal and transverse loss factor also can be used for describing the wakefields in the vacuum chamber, which can be calculated from the energy loss of the structure:[5,6]

$$k = \frac{\square\varepsilon}{q^2} \quad (3)$$

The (longitudinal) loss factor obtained by the integral over the longitudinal impedance $Z_{\square}(\omega)$ and the line density of a Gaussian distribution, and the transverse loss factor is also called kick factor, whose definition is analogous to that of the longitudinal loss factor:

$$\begin{cases} k_{\square}(\sigma) = \dfrac{c}{\pi}\displaystyle\int_0^{\infty} dk\, \mathrm{Re}\left(Z_{\square}(k)\right)\exp\left(-k^2\sigma^2\right) \\ k_{\perp}(\sigma) = -\dfrac{c}{\pi}\displaystyle\int_0^{\infty} dk\, \mathrm{Im}\left(Z_{\perp}(k)\right)\exp\left(-k^2\sigma^2\right) \end{cases} \quad (4)$$

## 3　Wake fields and impedances for HLS-II storage ring

In HLS-Ⅱ storage ring, we have a long electron bunch, so we can ignore the Coherent Synchrotron Radiation Wake Field and Impedance[7]; we also can ignore the influence of space charge effect, because of the ultra–relativistic particles[8]. So, in this paper, we just consider the resistive wall and the discontinuities wake fields,

### 3.1 Resistive wall

The vacuum chamber's octagonal cross-section of HLS-Ⅱ storage ring was shown in Fig 1, including its inner dimensions. Almost all of the chamber is made of Stainless Steel (SS).

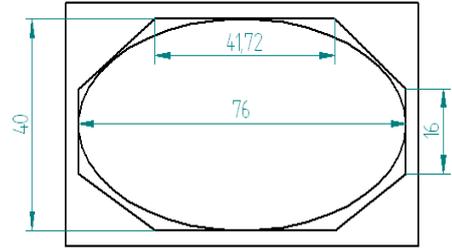

Fig.1. Cross-section of the SS vacuum chamber, dimensions are in unit of mm. A pipe indicated by an ellipse was used in the impedance estimation.

The impedance of the resistive wall is calculated analytically. In order to estimate the resistive wall impedance, we introduce a simplified modal(a simplified modal was introduced). Here all the cross section of the vacuum chamber is considered as elliptic, which has a length of major axis $2a$ and minor axis $2b$ showing in Fig 1.

$$a = d\,\cosh u_0, \quad b = d\,\sinh u_0, \quad d^2 = a^2 + b^2 \quad (5)$$

The longitudinal resistive wall impedances of this elliptic pipe are given by: [9,10]

$$\frac{Z_{\square}}{n} = Z_0 \frac{(1-i)}{2}\frac{\delta}{b}\frac{L}{2\pi R}G_0(u_0) \quad (6)$$

Where

$$G_0(u_0) = \frac{\sinh u_0}{2\pi}\int_0^{2\pi}\frac{Q_0^2(\upsilon)d\upsilon}{\left[\sinh^2 u_0 + \sin^2\upsilon\right]^{1/2}} \quad (7)$$

$$Q_0(\upsilon) = 1 + 2\sum_{m=1}^{\infty}(-1)^m\frac{\cos 2m\upsilon}{\cosh 2m u_0}$$

The transverse resistive wall impedance of an elliptic pipe can be estimated as:

$$Z_{\perp} = \frac{2R}{b^2}\frac{Z_{\square}}{n} \quad (8)$$

In this case, the conductivity $\sigma$ of 56m SS beam pipe is considered to be $1.5\times10^6\,(\Omega m)^{-1}$ and the permeability $\mu_0$ of free space is $4\pi\times10^{-7}\,\mathrm{H/m}$, then we

can obtain the impedance of the resistive wall with revolution frequency $f = 4.5334\text{MHz}$ :

$$\begin{cases} \dfrac{Z_\square}{n} = 4.5212(1-\text{i})\big/\sqrt{n}\ \Omega \\ Z_x = 33.865(1-\text{i})\big/\sqrt{n}\ \text{K}\Omega \\ Z_y = 36.8767(1-\text{i})\big/\sqrt{n}\ \text{K}\Omega \end{cases} \qquad (9)$$

### 3.2 BPMs

There are 36 BPMs (Fig. 2) in HLS-Ⅱ storage ring. Because of the resonant modes in the buttons of the BPMs (Fig. 3), the impedance, loss factor and kick factor are effected by the BPM geometry obviously. To estimate the transverse and longitudinal impedance, the designed BPMs were modeled on the octagonal cross-section for HLS-Ⅱ.

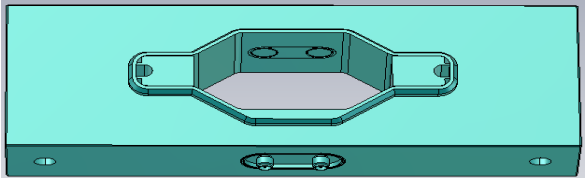

Fig.2. BPM

### 3.3 Coated ceramic vacuum chamber

In HLS-Ⅱ, there is a coated ceramic vacuum chamber. A mental coasting is required on the inner surface of the ceramic vacuum chamber, which is required to suppress penetration of the beam fields into the kicker at dangerous beam frequencies (about 1 MHz). For this kind of structure, it is very difficult to calculate the coupling impedance, because of the small depth of the metal coating. Hence a simple model of an infinitely long pipe was used to estimate the coating impedance.[11]

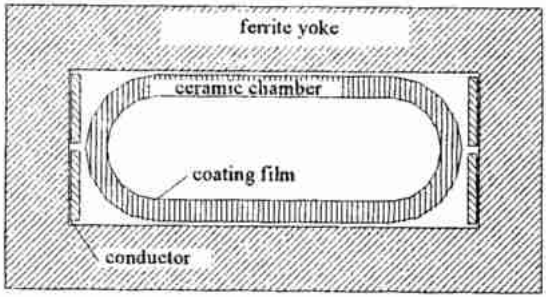

Fig.3. Cross section of ceramic vacuum chamber of injection system at HLS-Ⅱ

The longitudinal impedance from formula are presented below：

$$\frac{Z_\square}{L} = Z_{met}\frac{A + \tanh(\kappa d_m)}{1 + \tanh(\kappa d_m)} \qquad (10)$$

Where

$$Z_{met} = \frac{1 - i\,\text{sgn}(\omega)}{bc}\sqrt{\frac{|\omega|}{2\pi\sigma}}\frac{1}{\sqrt{4\pi\varepsilon_0}} \qquad (11)$$

$$A = \frac{\big(1 - i\,\text{sgn}(\omega)\big)d_c}{d_s}\left(1 - \frac{1}{\varepsilon}\right) \qquad (12)$$

$$\kappa = \frac{1 - i\,\text{sgn}(\omega)}{\delta} \qquad (13)$$

$$d_s = \frac{c}{\sqrt{2\pi\sigma|\omega|}}\sqrt{4\pi\varepsilon_0} \qquad (14)$$

In HLS-Ⅱ, the length of this piece $L$ is 0.28m, the thickness of the metal layer $d_m$ is $10^{-6}$m, the thickness of the ceramic chamber $d_c$ is 0.005m, the equivalent radius of the vacuum chamber is 0.2m, we can obtain that: $Z_\square = 0.002497n - 0.014774\text{i}$ .

### 3.4 Pump ports

In present HLS-Ⅱ storage ring, there are about 20 vacuum Pumps, the ports of them are screened with some small holes, which is composed by a rectangle and two half circles. Because of the different arrange of the holes and water-cooled, there are five kinds of pump ports, and their longitudinal and transverse impedance, loss factor, kick factor were calculated by CST code.

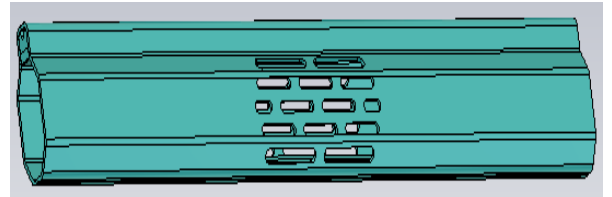

Fig.4. Pumping port

### 3.5 Longitudinal feedback kicker

The longitudinal feedback kicker (Fig.3) installed in the injection devices of the upgrading HLS-Ⅱ storage ring is a waveguide overloaded pillbox cavity with two input and two output ports, and the pillbox gap is 94mm. The dimension of the HLS-Ⅱ LFB kicker has the similar performance as the Duke kicker, which is built in a tapered section to transition from the round beam pipe to the octagonal vacuum chamber. This method leads to a strong coupling between the pillbox and the waveguides, which can help damp the harmful HOMs. But there is also a small disadvantage of the LFB kicker, that the broadband impedance of the accelerator is increased.

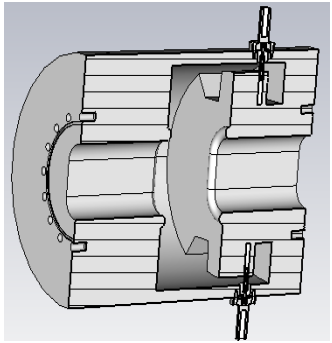

Fig.5. Longitudinal feedback kicker

### 3.6 Clearing electrodes

There are 32 clearing electrodes built in the HLS-Ⅱ storage ring, which have a length of 525 or 615mm as showing in Fig. 6. The electrodes have been inserted in the machine through ceramic material supports, which allowed inserting the electrons in the vacuum chamber without damping. Thus, the the full-height of the vacuum chamber is 43.5mm, which is 3.5mm higher than the common vacuum chamber.

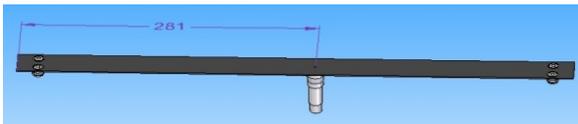

Fig.6. Clearing electrode

### 3.7 RF cavity

At present, there is one RF cavity (Fig.4) in the HLS-Ⅱ storage ring, which is important sources of impedance. The 3D model of the RF cavity is built by CST code, which is shown in Fig.7, its length is 0.472m. Only the fundamental mode of the cavity has high quality factor and hence high impedance, the other modes have low quality factors due to energy loss.

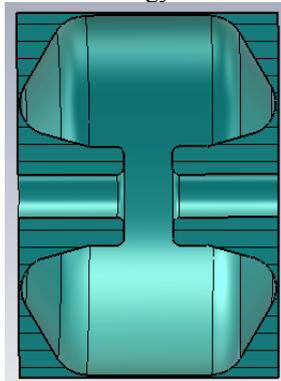

Fig.7. RF Cavity

### 3.8 Others

In the HLS-Ⅱ storage ring, there are a large number of small discontinuities in the vacuum chamber of different kinds and sizes, such as slots for synchrotron radiation monitor, steps between the BPMs and the vacuum chambers etc. Meanwhile we also calculated the wake fields and impedances about the curved section, flange, bellows and so on, which are not described in details.

## 4  Impedance budget

Each component of HLS-Ⅱ storage ring was simulated so far(has been simulated). Fig. 1(?) shows the wake fields of several kinds of elements as well as the total wake field. [5] Then we can find that the (the results suggested that the )most significant wake potentials are produced by pumping ports, clearing electrodes, longitudinal feedback kicker, RF cavity and curved section. Meanwhile, because of the similar cross section, there is little deference between each one. In the calculations , the transverse wake fields is much smaller than the longitudinal one, which have little influence of(on) the beam dynamics, so in this paper, we just list the longitudinal wake fields of each component of HLS-Ⅱ storage ring.

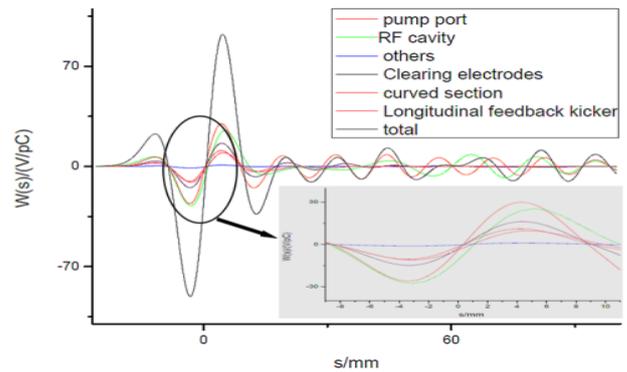

Fig1. Longitudinal wake fields as a function of distance from bunch for various groups of elements.

Impedance budget[12,13] of the HLS-Ⅱ storage ring is shown in Table 1, where present the longitudinal broadband impedance, longitudinal loss factor, and the transverse kick factor are element of each elements. The major contributors to the impedance of the storage ring are RF cavity, pumping port chamber, longitudinal feedback kicker and flanges. In Table 2, the broadband impedance, loss factor and kick factor are the sum of each kind of components in the HLS-Ⅱ storage ring. The results indicated that the total longitudinal broadband impedance is about1.08 Ω , the total longitudinal loss factor is $27.7V/pC$ .

Then the broadband impedance model for storage ring is obtained as:

Table 2 :calculated impedance for components of the HLS-II storage ring

| Component | $N$ | $\left|Z_0/n\right|$ $\Omega$ | $k$ V/pC | $k_x$ V/pC/mm | $k_y$ V/pC/mm |
|---|---|---|---|---|---|
| RF cavity | 1 | 0.041032 | 3.65218 | 6.6289e-1 | 6.6289e-1 |
| Pumping port | 25 | 0.068769 | 1.5672 | 3.9795e-2 | 2.9964e-2 |
| Flange | 59 | 0.061478 | 1.7994 | 3.4842e-3 | 2.0060e-3 |
| BPM | 35 | 0.047419 | 3.993 | 2.0763e-3 | 8.6992e-4 |
| Clearing electrodes | 32 | 0.091903 | 6.2462 | 1.4910e-1 | 1.6569e-1 |
| bellows | 16 | 0.017048 | 0.25872 | 9.8884e-3 | 8.9129e-3 |
| Longitudinal feedback kicker | 1 | 0.08444 | 1.31425 | 1.8352e-1 | 1.9410e-1 |
| Resistive wall | - | 0.6740 | 4.283 | 1.6289e-1` | 1.7582e-1 |
| Others | - | 0.00153 | 4.57816 | 2.1344e-1 | 2.4100e-1 |
| Total | 169 | 1.087619 | 27.69211 | 1.43E+00 | 1.48E+00 |

$$Z_0/n = 1.087619i + \frac{4.5212(1-i)}{\sqrt{n}} + 0.002497n \quad (15)$$

Where the first term of the formula is the broad-band impedance at low frequency, the second term is the resistive wall impedance, the last term is the impedance of the coated ceramic vacuum chamber.

## 5 Instability threshold estimation

The Total value of the longitudinal broad-band impedance was used to study the longitudinal microwave instabilities in the HLS-II storage ring.

### 5.1 Parasitic energy loss

When a beam traverses in the storage ring, it will loses a certain amount of energy to the impedance，which we are called parasitic loss. In HLS-II storage ring, the broadband impedance was used to get the parasitic loss power:[3]

$$p_{parasitic} = -\frac{\Box\varepsilon}{T_0} = -\frac{kq^2}{T_0} \quad (16)$$

Where $k$ is the loss factor, $q$ is the charge of the bunch, $T_0$ is the revolution period of the beam around the accelerator. In the HLS-II storage ring, the parasitic loss power is 3.197eV.

### 5.2 Microwave Instability

When the peak current of a single bunch is higher than a threshold current, the bunch lengthening and energy spread are generated until the peak current is reduced to the current threshold, meanwhile, the microwave instabilities [14] also affect the brightness of the insertion device.

$$I_{pk} = \frac{\sqrt{2\pi}R}{\sigma_l}I_b = \frac{2\pi\alpha(E/e)\delta^2}{\left|Z_0/n\right|} \quad (17)$$

Where $I_{pk}$ is the peak current threshold, $\sigma_l$ is the rms bunch length, $\delta$ is the rms fractional spread. In the HLS-II storage ring, the threshold peak current of the microwave instability is 2.059mA.

If we apply the Keil-Schnell criterion, we can obtain the bunch lengthening relation due to only the microwave instability as given by:

$$\sigma_l^3 = \frac{\alpha c^2 R\left|Z_0/n\right|}{\sqrt{2\pi}(E/e)v_s^2\omega_0^2}I_b \quad (18)$$

Where $v_s \approx \sqrt{\frac{ehV_{rf}\alpha}{2\pi E_0}}$ is the synchrotron turn, $\omega_0$ is revolution frequency, $h$ is the harmonic number, $V_{rf}$ is the rf voltage. So we can obtain the length of the lengthened bunch is 128ps.

### 5.3 Transverse mode coupling instability

When the beam current is increased, the amplitude of some azimuthal modes may be larger than those of the betatron fractional turns, and then, the frequency will be changed, the beams will be lost. In this case, some of the azimuthal modes will coupled, which can not be distinguished, which is called transverse mode coupling instability[15]. The transverse mode-coupling instability is driven by the imaginary part of the transverse impedance and its threshold is given by:

$$I_T^{th} = \frac{4(E/e)v_s}{Z_T\langle\beta_T\rangle R}b \quad (19)$$

Where $\langle\beta_T\rangle$ is the ring-average transverse beta function and $R$ is the ring average radius. Use the numerical value in the Table 1, we can obtain the single bunch current threshold is 83.6mA, which is much larger than the one of longitudinal microwave instabilities. So the longitudinal microwave instabilities are much significant in the storage ring.

# 6 Conclusions

Using analytical and numerical methods，we calculated the short-range wakefields and broadband impedances of the HLS-Ⅱ storage ring, contributions to the total longitudinal loss factor, transverse kick factor and total longitudinal broadband impedance. Meanwhile we estimate the stabilities based the broadband impedance, which did not affect the beam quality seriously. In the later studies, we will use the broadband impedance and the narrowband impedance to study the beam instabilities by tracking.

## Reference:

[1] Li Wei-min, Xu Hong-liang, Wang lin, Feng Guang-yao, Zhang Shan-cai, Hao Hao, "The Upgraded scheme of Hefei Light Source ", AIP Conf. Proc.

[2] Zhang Shancai, Li Weimin, Feng Guangyao, Wang Lin, Gao Weiwei, Xu Hongliang, Fan Wei, "THE UPGTADE OF HEFEI LIGHT SOURCE(HLS) TRANSPORT LINE", Proceeding of IPAC'10, Kyoto, Japan.

[3] Chao A．Physics of Collective Beam Instabilities in High Energy Accelerators．New York：Wiley Inter science．19933．

[4] K. Y. Ng, Physics of Intensity Dependent Beam Instabilities (World Scientific Publishing Co. Pte. Ltd. 2006).

[5] B. Zotter and S. Kheifets, Impedances and Wakes in High Energy Particle Accelerators. London: W World Scientific, 1998.

[6] R. Bartolini, R. Fielder, C. Thomas, LOSS FACTOR AND IMPEDANCE ANALYSIS FOR THE DIAMOND STORAGE RING. Proceeding of IPAC'13, Shanghai, China.

[7] S. A. Kheifets, B. Zotter, the Coherent Synchrotron Radiation Wake Field and Impedance. CERN SL Report 95-43 (AP).

[8] G. Stupakov and Z. Huang, Space charge effect in an accelerated beam. SLAC-PUB-12768, August 2007

[9] R. L. Gluckstern, J. van Zeijts, B. Zotter, Coupling Impedance of Beam Pipes of General Cross Section, CERN SL/AP 92-25, June 1992.

[10] J. Byrd, G. Lambertson Resistive wall impedance of the B-factory, PEP-II AP Note 9-93, March 1993

[11] Danilov V. An Improved Impedance Model of Metallic Coatings, EPAC 2002, 1464-1466.

[12] S.Y.Zhang,"SNS Storage Ring Impedance", BNL/SNS TECHNICAL NOTE

[13] A. Blednykh, S. Krinsky and J. Rose, "Coupling impedance of CESR-B RF cavity for the NSLS-II storage ring" PAC 2007.

[14] C. Kim, Yujong Kim, K. H. Kim, J. Y. Huang, M. H. Cho, W. Namkung, and I. S. KO. SINGLE BUNCH COLLECTIVE EFFECTS IN PLS STORAGE RING. PAC 2002, Chicago.

[15] M. P. Level et al., "Transverse Mode-Coupling Experiment in DCI," LAL/RT/84-09, 1984.